\documentclass{article}
\pdfoutput=1
\usepackage{spconf,amsmath,graphicx}

\usepackage{mwe} 

\def\b{\ensuremath\boldsymbol}

\usepackage{amsfonts}
\usepackage{xcolor}
\usepackage{subcaption}  

\usepackage{atbegshi,picture}

\title{Magnification Generalization for Histopathology Image Embedding}
\name{Milad Sikaroudi$^{\star \ddagger}$, Benyamin Ghojogh$^{\dagger \ddagger}$, Fakhri Karray$^{\dagger}$, Mark Crowley$^{\dagger}$, H.R. Tizhoosh$^{\star}$\thanks{$\ddagger$ The first two authors contributed equally to this work.}}

\address{$^{\star}$KIMIA Lab, University of Waterloo, ON, Canada\\
    $^{\dagger}$Department of Electrical and Computer Engineering, University of Waterloo, ON, Canada}

\AtBeginShipout{\AtBeginShipoutUpperLeft{%
  \put(\dimexpr\paperwidth-1cm\relax,-1.5cm){\makebox[0pt][r]{\framebox{\small Accepted for presentation at  International Symposium on Biomedical Imaging (ISBI'2021)}}}%
}}

\begin{document}

%
\maketitle
\begin{abstract}
Histopathology image embedding is an active research area in computer vision. Most of the embedding models exclusively concentrate on a specific magnification level. However, a useful task in histopathology embedding is to train an embedding space regardless of the magnification level. Two main approaches for tackling this goal are domain adaptation and domain generalization, where the target magnification levels may or may not be introduced to the model in training, respectively. Although magnification adaptation is a well-studied topic in the literature, this paper, to the best of our knowledge, is the first work on magnification generalization for histopathology image embedding. We use an episodic trainable domain generalization technique for magnification generalization, namely Model Agnostic Learning of Semantic Features (MASF), which works based on the Model Agnostic Meta-Learning (MAML) concept. Our experimental results on a breast cancer histopathology dataset with four different magnification levels show the proposed method's effectiveness for magnification generalization.
\end{abstract}
\begin{keywords}
Histopathology, magnification, domain generalization, model agnostic, semantic features
\end{keywords}

\section{Introduction}

Recently, machine learning and particularly deep learning have been widely utilized for cancer diagnosis in the histopathology domain \cite{kalra2020pan}. Due to the large size of the images, the literature mostly tries to learn a discriminative embedding space for histopathology image patches (i.e., sub-images). A promising embedding space imposes variance between the classes of histopathology tissue or cancer types while reducing the intra-class scatters \cite{ghojogh2020fisher}. 

Many of the embedding methods learn the subspace for only a specific magnification. However, one of the main challenges in histopathology image embedding is the different magnification levels for indexing of a Whole Slide Indexing (WSI) image \cite{sellaro2013relationship}. It is well-known that significantly different patterns may exist at different magnification levels of a WSI  \cite{zaveri2020recognizing}. 
It is useful to train an embedding space for discriminating the histopathology patches regardless of their magnifications. That would lead to learning more compact WSI representations. It has been an arduous task because of the significant domain shifts between different magnification levels with noticeably different patterns.

There are generally two approaches to learn a discriminative embedding space regardless of magnification level. One approach is called ``domain adaptation'' \cite{ben2010theory}, in which all different magnification levels in the training phase are used to push the embeddings with the same class but different magnification levels toward each other. By doing so, the same-class patches in the embedding space fall close to each other regardless of their magnification levels. Another approach is ``domain generalization'' \cite{dou2019domain}, where the target domain, to which we intend to generalize, has not been observed in the training phase; hence, the embedding space is learned using other source domains. Domain generalization is a more challenging task than domain adaptation because the target magnification level is not introduced to the model. Note that domain generalization is more realistic than domain adaptation because in many cases new domain shifts, which were not available in the training phase, are introduced in the test phase \cite{dou2019domain}. 

There exist several works on magnification adaptation for histopathology embedding in the literature \cite{bayramoglu2016deep,das2017classifying,spanhol2016breast}. These works train an embedding space for discriminating the histopathology patches using all available magnification levels. However, to the best of our knowledge, this paper is the first work on magnification generalization for histopathology image embedding where the target magnification is not introduced to the model in training. We use a recently proposed domain generalization method, named Model Agnostic learning of Semantic Features (MASF) \cite{dou2019domain} which is based upon the Model Agnostic Meta-Learning (MAML) \cite{finn2017model}. The proposed magnification generalization can demonstrate how generalizable is each magnification level based on the rest of the magnification levels. This magnification generalization can be conducive to provide a more compact representation for WSI images.
The remainder of this paper is organized as follows: We review the MAML and MASF models in Section \ref{section_review}. The proposed magnification generalization for histopathology image embedding is detailed in Section \ref{section_methodology}. The experimental results are reported and discussed in Section \ref{section_experiments}. Finally, Section \ref{section_conclusion} concludes the paper. 

\section{Literature Review}\label{section_review}

In this section, we review MAML which has inspired the MASF method. Afterward, we introduce MASF which we use for magnification generalization in histopathology image embedding. 

\subsection{Model Agnostic Meta Learning}

MAML \cite{finn2017model} is a method proposed for meta-learning. It can be used for either few-shot learning \cite{vanschoren2018meta} or domain generalization \cite{dou2019domain}. 
It updates a gradient through another gradient. Suppose we have $K$ training source domains, denoted by $\{\mathcal{D}_k\}_{k=1}^K$, where the learned embedding space is supposed to be generalizable to other unseen target domains. Let the learning model have learnable parameters $\theta$. We denote the model with parameter $\theta$ by $F_\theta$. The training phase of MAML is iterative wherein every iteration, some training samples are considered from every domain. 
MAML applies a step of gradient descent for every training domain:
\begin{align}\label{equation_MAML_update}
\theta'_i := \theta - \rho \nabla_\theta \mathcal{L}(\mathcal{D}_k; \theta),
\end{align}
where $\rho$ is the learning rate, $\nabla_\theta$ denotes derivative with respect to $\theta$ and $\mathcal{L}({\mathcal{D}_k}; \theta)$ is the loss function over the embedding of samples of the $k$-th domain using parameter $\theta$. After this step, we have the parameters  $\{\theta'_k\}_{k=1}^K$. Finally, the meta update is done using a step of gradient descent:
\begin{align}\label{equation_MAML_meta_update}
\theta := \theta - \sigma \nabla_\theta \sum_{k=1}^K \mathcal{L}(\mathcal{D}_k; \theta'_k),
\end{align}
where $\sigma$ is the meta learning rate and $\mathcal{L}({\mathcal{D}_k}; \theta'_k)$ is the loss function over the embedding of samples of the $k$-th domain using parameter $\theta'_k$. 
The Eqs. (\ref{equation_MAML_update}) and (\ref{equation_MAML_meta_update}) are repeated iteratively until convergence.

\subsection{Model Agnostic Semantic Features}

\begin{figure}[!t]
\centering
\includegraphics[trim=0 0 50 0,clip,width=3.1in]{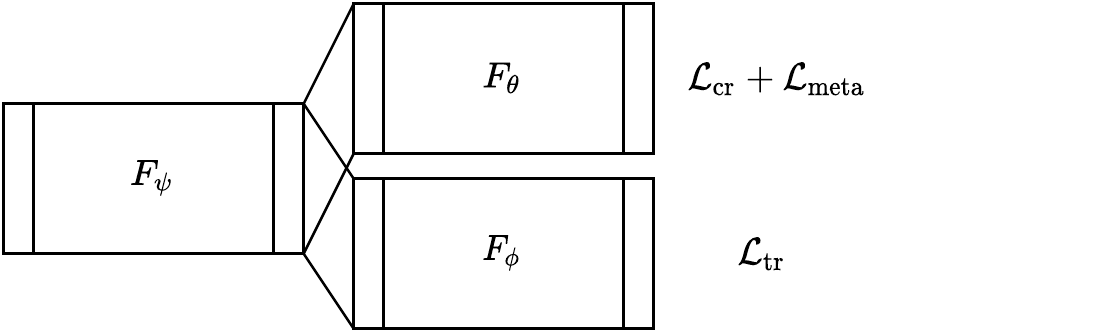}
\caption{The subnetworks and loss functions used in MASF and the proposed magnification generalization for histopathology image embedding.}
\label{figure_losses}
\end{figure}

MASF \cite{dou2019domain} is a meta-learning method for domain generalization. Inspired by the MAML approach, it learns an embedding space suitable for generalization to an unseen target domain. 
The MASF method includes a neural network for feature extraction, some layers for classification with cross-entropy loss, and some layers for metric embedding, denoted by $F_\psi$, $F_\theta$, and $F_\phi$, respectively, with the parameters (weights) $\psi$, $\theta$, and $\phi$.
These subnetworks are used for the three different loss functions of which the MASF training is composed. The subnetworks $F_\theta$ and $F_\phi$ both come after the sub-network $F_\psi$, as illustrated in Fig. \ref{figure_losses}. 

MASF combines three different loss functions for learning a discriminative embedding space. These loss functions are a cross-entropy loss for supervised embedding, soft confusion matrix for domain generalization, and a triplet loss for metric learning. 
In this work, we use the MASF method for magnification generalization in histopathology image embedding.

\section{Magnification Generalization}\label{section_methodology}

Consider the $K$ training source magnifications, $\{\mathcal{M}_k\}_{k=1}^K$. In every iteration of the MASF method, the source magnifications are split into meta-train and meta-test magnifications, denoted by $\mathcal{M}_\text{tr}$ and $\mathcal{M}_\text{te}$, respectively. 
As in MASF, we use a combination of three different loss functions for learning the embedding space. These loss functions are detailed in the following.

\subsection{Loss for Supervised Embedding}

The first loss which we use is the cross-entropy loss for supervised embedding: 
\begin{align}
&\mathcal{L}_\text{cr}(\mathcal{M}_\text{tr}; \psi, \theta) := \nonumber \\
&\frac{-1}{|\mathcal{M}_\text{tr}|} \sum_{\mathcal{M} \in \mathcal{M}_\text{tr}} \frac{1}{|\mathcal{M}|} \sum_{(\b{x},y) \in \mathcal{M}} \sum_{c=1}^C \mathbb{I}(y = c) \log \mathbb{I}(\widehat{y} = c), \label{equation_cross_entropy}
\end{align}
where $C$ is the number of classes, $y$ is the ground truth label for $\b{x}$, $\widehat{y}$ is the predicted label for $\b{x}$, and $\mathbb{I}(\cdot)$ is the indicator function which is one when its condition is satisfied and is zero otherwise. 
In practice, we use Eq. (\ref{equation_cross_entropy}) in batches.
We denote the updated $\psi$ and $\phi$ by the cross-entropy loss by $\psi'$ and $\phi'$, respectively. 

\subsection{Loss for Magnification Generalization}

The next loss used is the generalization loss which takes care of magnification generalization. 
Consider the feature extraction embedding of samples of magnification $k$. If the samples of this magnification in the class label $c$ are denoted by $\{\b{x}_{c,i}^{(k)}\}_{i=1}^{n_c}$ and their feature extraction embeddings are $\{F_{\psi'}(\b{x}_{c,i}^{(k)})\}_{i=1}^{n_c}$, the Monte-Carlo approximation for the mean embedding of class $c$ in magnification $k$ can be given as
\begin{align}\label{equation_z}
\b{z}_c^{(k)} := \frac{1}{n_c} \sum_{i=1}^{n_c} F_{\psi'}(\b{x}_{c,i}^{(k)}).
\end{align}
This mean is passed through the layers for supervised embedding and its softmax with temperature $\tau > 1$ is
\begin{align}
\b{s}_c^{(k)} := \text{softmax}(F_{\theta'}(\b{z}_c^{(k)})/\tau),
\end{align}
which provides a soft confusion matrix.

For two magnifications $\mathcal{M}_i$ and $\mathcal{M}_j$, the global loss, the symmetrized Kullback–Leibler (KL) divergence, averaged over all the $C$ classes, is calculated as
\begin{align}
&\ell_\text{gen}(\mathcal{M}_i, \mathcal{M}_j; \psi', \theta') \nonumber \\
&~~ := \frac{1}{C} \sum_{c=1}^C \frac{1}{2} \Big[ D_\text{KL}(\b{s}_c^{(i)} \| \b{s}_c^{(j)}) + D_\text{KL}(\b{s}_c^{(j)} \| \b{s}_c^{(i)}) \Big], \label{equation_l_gen}
\end{align}
where $D_\text{KL}$ denotes the KL divergence. 
This loss is computed over all the meta-train and meta-test magnifications through 
\begin{align}
& \mathcal{L}_\text{gen}(\mathcal{M}_\text{tr}, \mathcal{M}_\text{te}; \psi', \theta') := \nonumber \\
& \frac{1}{|\mathcal{M}_\text{tr}| |\mathcal{M}_\text{te}|}  \sum_{\mathcal{M}_i \in \mathcal{M}_\text{tr}} \sum_{\mathcal{M}_j \in \mathcal{M}_\text{te}} \ell_\text{gen} (\mathcal{M}_i, \mathcal{M}_j; \psi', \theta'), \label{equation_L_gen}
\end{align}
where $|\cdot|$ denotes the cardinality of set. 
In practice, we use Eqs. (\ref{equation_z}), (\ref{equation_l_gen}), and (\ref{equation_L_gen}) on  batches.

\subsection{Loss for Metric Learning}

Finally, a triplet loss \cite{schroff2015facenet} is used for the sake of metric learning. Consider a triplet of anchor, positive, and negative instances denoted by $\b{x}_a$, $\b{x}_p$, and $\b{x}_n$, respectively. Triplet loss attempts to increase the inter-class variances of data and decrease the intra-class variances. 
If we randomly sample $R$ triplets, $\mathcal{T} := \{\b{x}_a^r, \b{x}_p^r, \b{x}_n^r\}_{r=1}^R$, from all the source magnifications $\{\mathcal{M}_k\}_{k=1}^K$, the average triplet loss is 
\begin{align}
&\mathcal{L}_\text{tri}(\mathcal{T}; \psi', \phi') := \nonumber \\
&\frac{1}{R} \sum_{r=1}^R \Big[\|F_{\phi'}(F_{\psi'}(\b{x}_a^r)) - F_{\phi'}(F_{\psi'}(\b{x}_p^r))\|_2^2 \nonumber \\
&~~~~~~ + \|F_{\phi'}(F_{\psi'}(\b{x}_a^r)) - F_{\phi'}(F_{\psi'}(\b{x}_n^r))\|_2^2 + \zeta\Big]_+,
\end{align}
where $\|\cdot\|_2$ is the $\ell_2$ norm, $\zeta$ is a margin and $[\cdot]_+ := \max(\cdot,0)$ is the standard Hinge loss. 

\subsection{Updating Parameters}

Here, we explain how the weights of sub-networks, shown in Fig. \ref{figure_losses}, are updated iteratively. 
A step of gradient descent is used for updating the weights $\psi$ and $\theta$ given by
\begin{align}\label{equation_gradient_descent_cross_entropy}
(\psi', \theta') := (\psi, \theta) - \alpha \nabla_{\psi, \theta} \mathcal{L}_\text{cr}(\mathcal{M}_\text{tr}; \psi, \theta),
\end{align}
where $\alpha$ is the learning rate and $\nabla_{\psi, \theta}$ denotes gradient with respect to $\psi$ and $\theta$. 

A linear combination of the generalization and triplet losses is used for the meta loss:
\begin{align}
&\mathcal{L}_\text{meta}(\mathcal{M}_\text{tr}, \mathcal{M}_\text{te}, \mathcal{T}; \psi', \theta', \phi') := \nonumber \\
&~~~~~~~~ \beta_1 \mathcal{L}_\text{gen}(\mathcal{M}_\text{tr}, \mathcal{M}_\text{te}; \psi', \theta') + \beta_2 \mathcal{L}_\text{tri}(\mathcal{T}; \psi', \phi'),
\end{align}
where $\beta_1, \beta_2 > 0$. 
After Eq. (\ref{equation_gradient_descent_cross_entropy}), the weights are updated using two other gradient descent steps:
\begin{align}
& (\psi, \theta) := (\psi, \theta) - \eta \nabla_{\psi, \theta} (\mathcal{L}_\text{cr} + \mathcal{L}_\text{meta}), \label{equation_gradient_descent_cross_entropy_and_meta} \\
&\phi := \phi - \gamma \nabla_\phi \mathcal{L}_\text{tri}(\mathcal{T}; \psi', \phi'), \label{equation_gradient_descent_triplet_loss}
\end{align}
where $\eta$ and $\gamma$ are the learning rates. 
Eqs. (\ref{equation_gradient_descent_cross_entropy}), (\ref{equation_gradient_descent_cross_entropy_and_meta}), and (\ref{equation_gradient_descent_triplet_loss}) are repeated iteratively until convergence. These equations are performed using the backpropagation algorithm. Figure \ref{figure_losses} shows these loss functions.

\section{Experimental Results}\label{section_experiments}

\subsection{Dataset, Preprocessing, and Setup}

We use a breast cancer histopathology image dataset (BreaKHis) \cite{spanhol2015dataset} for evaluation. This dataset consists of four types of benign breast tumors, i.e, adenosis (A), fibroadenoma (F), phyllodes tumor (PT), and tubular adenoma (TA) as well as four malignant breast cancer tumors, i.e., ductal carcinoma (DC), lobular carcinoma (LC), mucinous carcinoma (MC), and papillary carcinoma (PC). The histopathology patches are provided in four different magnification levels, i.e., 40$\times$, 100$\times$, 200$\times$, and 400$\times$. 

As preprocessing, we applied the Reinhard stain normalization \cite{reinhard2001color} to the images. 
The dataset was split into training-validation-test sets with portions 45\%-45\%-10\% inspired by the PCAS dataset. 
We used AlexNet \cite{krizhevsky2012imagenet} as the backbone and the initialized weights with the pre-trained ImageNet weights \cite{russakovsky2015imagenet}. 
For experiments, we set $\beta_1 = 1, \beta_2 = 0.005, \alpha = \eta = \gamma = 10^{-5}$ inspired by \cite{dou2019domain}.
For the remaining network configuration and hyper-parameters, we used the same setup as in \cite{dou2019domain} with the PACS dataset. We used the Adam optimizer and early stopping for avoiding overfitting. 

\subsection{Magnification Generalization for Tumor Types}

Inspired by literature \cite{dou2019domain}, our baseline for comparison of generalization is the DeepAll approach in which all the source domains are combined to train an embedding space with a cross-entropy loss function. 
Table \ref{table_tumore_types} reports the average accuracy of embeddings by our and DeepAll models where the accuracies are averaged over three independent runs, inspired by \cite{dou2019domain}. In this experiment, all eight tumor types are used as classes which is a demanding task due to the presence of similarity of patterns between different tumor types.
Four different cases are reported where we leave one domain out of the magnification levels to be considered as the target magnification. 
This generalization is useful in practice for cancer diagnosis of novel magnifications. 
The table shows we outperform the baseline in all cases of target magnifications. 

\begin{table}[!t]
\caption{Comparison of magnification generalization for tumor types by DeepAll and our method based on MASF. The rates are accuracy percent.}
\label{table_tumore_types}
\centering
\scalebox{0.86}{    
\begin{tabular}{c || c | c | c }
Source & Target & Ours & DeepAll   \\
\hline
\hline
100$\times$, 200$\times$, 400$\times$ & 40$\times$ & 44.37 $\pm$ 0.11 & 40.39 $\pm$ 0.46 \\
40$\times$, 200$\times$, 400$\times$ & 100$\times$ & 50.82 $\pm$ 0.06 & 48.82 $\pm$ 0.39 \\
40$\times$, 100$\times$, 400$\times$ & 200$\times$ & 52.82 $\pm$ 0.22 & 51.75 $\pm$ 0.26 \\
40$\times$, 100$\times$, 200$\times$ & 400$\times$ & 48.27 $\pm$ 0.43 & 47.21 $\pm$ 0.19 \\
\hline
\hline
\end{tabular}%
}
\end{table}

\subsection{Magnification Generalization for Malignancy}

We also evaluate our proposed method for magnification generalization on the binary classification on the malignancy status, i.e., benign or malignant. 
For this experiment, the average accuracy of embeddings by our and DeepAll models over three independent runs are reported in Table \ref{table_malignancy}. 
As binary classification is easier than the previous experiment, the accuracies are higher in this table. Note that the rates in this table are slightly lower than those in the magnification adaptation methods \cite{bayramoglu2016deep,das2017classifying,spanhol2016breast} because generalization is a more arduous task than adaptation \cite{dou2019domain}. 
As the table reports, we outperform the baseline in all cases of target magnifications demonstrating the effectiveness of the proposed method. 

\begin{figure}[!t]
\centering
\begin{subfigure}[b]{0.49\textwidth}
\centering
\includegraphics[width=1.7in]{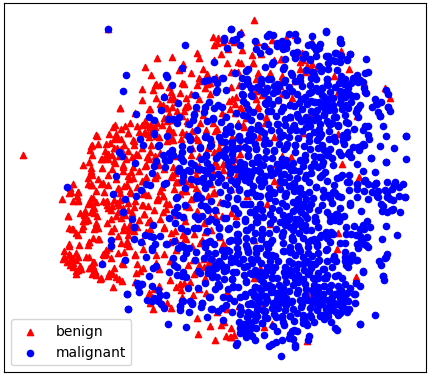} 
\caption{Classification of embedding w.r.t. malignancy.}
\label{figure_malignancy}
\end{subfigure}
\begin{subfigure}[b]{0.49\textwidth}
\centering
\includegraphics[width=1.7in]{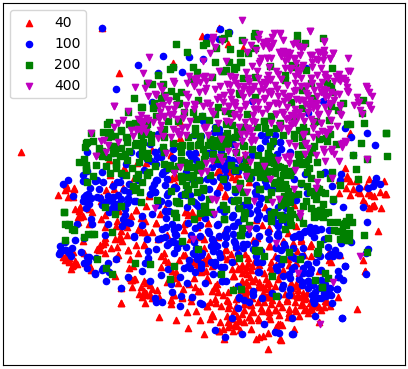} 
\caption{Classification of embedding w.r.t. magnification levels.}
\label{figure_magnification}
\end{subfigure}
\caption{Embedding of the test dataset with the target magnification $400\times$.}
\label{figure_embedding}
\end{figure}

Figure \ref{figure_embedding} shows an example of trained embedding, with target magnification $400\times$ where the embedding is plotted with both malignancy and magnification labels. In either case, the classes including the unseen target magnification are well separated. 
Moreover, it can be seen that both benign and malignant instances include all different magnification levels, even the $400\times$ magnification level unseen during training. 
The meaningfulness of the trained embedding space also substantiates the effectiveness of the proposed magnification generalization. 

\begin{table}[!t]
\caption{Comparison of magnification generalization for malignancy by DeepAll and our method based on MASF. The rates are accuracy percent.}
\label{table_malignancy}
\centering
\scalebox{0.86}{    
\begin{tabular}{c || c | c | c }
Source & Target & Ours & DeepAll  \\
\hline
\hline
100$\times$, 200$\times$, 400$\times$ & 40$\times$ & 81.01 $\pm$ 0.42 & 78.86 $\pm$ 0.23 \\
40$\times$, 200$\times$, 400$\times$ & 100$\times$ & 80.98 $\pm$ 0.70 & 80.52 $\pm$ 0.34 \\
40$\times$, 100$\times$, 400$\times$ & 200$\times$ & 80.99 $\pm$ 0.57 & 79.96 $\pm$ 0.86 \\
40$\times$, 100$\times$, 200$\times$ & 400$\times$ & 77.25 $\pm$ 0.83 & 74.44 $\pm$ 0.31 \\
\hline
\hline
\end{tabular}%
}
\end{table}

\section{Conclusion and Future Direction}\label{section_conclusion}

In this paper, we proposed magnification generalization for histopathology image embedding where the target magnification level is not introduced to the model in the training phase. The proposed model uses MASF which has been based upon MAML tailored to perform domain generalization. Three loss functions, i.e., the cross-entropy, softmax, and triplet loss, are used for learning the embedding space. Experimental results on the BreaKHis dataset verified the effectiveness of this generalization.  A possible future direction is to incorporate more discriminative metric embedding \cite{do2019theoretically}, \cite{ghojogh2020fisher}. 

\subsection*{Acknowledgments}
We would like to thank the Ontario Government for funding our ORF-RE consortium on the identification of histopathology gigapixel images. The authors declare no conflict of interest.  

\subsection*{Compliance with Ethical Standards}
 This paper has compliance with ethical standards.


\bibliographystyle{IEEEbib}
\bibliography{refs}

\end{document}